%% file: main.tex
\documentclass[conference]{IEEEtran}
\usepackage{url}
\usepackage{mathtools} 
\usepackage{graphicx}
\usepackage{bbm}
\usepackage{algorithm}
\usepackage{amsfonts}
\graphicspath{{figures/}}
\usepackage{multirow}
\usepackage{subfigure}

\usepackage{algcompatible}
\usepackage[dvipsnames]{xcolor}
\pdfoutput=1
\include{my_definitions}

\newcommand{\pushcode}[1][1]{\hskip\dimexpr#1\algorithmicindent\relax}

\newcommand{\STATExx}{\STATEx\hspace{-1pt}}
\usepackage{forloop}
\newcounter{loopcntr}
\newcommand{\rpt}[2][1]{%
              \forloop{loopcntr}{0}{\value{loopcntr}<#1}{#2}%
}

\newcommand{\subgroup}[1]%
{\rlap{\smash{%
              \newcount\cnt%
              \cnt \numexpr#1\relax%
              \advance\cnt -1\relax%
              $\tabcolsep=.2em\begin{tabular}[t]{|l}\multicolumn{1}{l}{}\\%
              \rpt[\cnt]{\\}
              \\\hline\end{tabular}$%
}}}

\newcounter{myRefCount}

\begin{document}

\title{Deep Learning Assisted CSI Estimation for Joint URLLC and eMBB Resource Allocation}

\author{
\IEEEauthorblockN{
    Hamza Khan\IEEEauthorrefmark{1},
    M. Majid Butt\IEEEauthorrefmark{2},
	Sumudu Samarakoon\IEEEauthorrefmark{1},
    Philippe Sehier\IEEEauthorrefmark{2},
	and Mehdi Bennis\IEEEauthorrefmark{1}
	\\}
\IEEEauthorblockA{
	\small%
	\IEEEauthorrefmark{1} Centre for Wireless Communication, University of Oulu, Finland, \\
	emails: \{hamza.khan,\,sumudu.samarakoon,\,mehdi.bennis\}@oulu.fi \\
	\IEEEauthorrefmark{2}Nokia Bell Labs, Paris-Saclay, France, \\ emails: \{majid.butt,\,philippe.sehier\}@nokia-bell-labs.com
}
\vspace{-0.6cm}}

\maketitle

\begin{abstract}
\input{abstract.tex}
\end{abstract}

\IEEEpeerreviewmaketitle
\section{Introduction}
\input{introduction.tex}

\section{Resource allocation of vehicular network under perfect CSI}
\input{system_model.tex}

\section{Resource allocation using CSI inference}
\input{CSI_inference.tex}

\section{Simulation Results}
\input{simulation.tex}

\section{Conclusion}
\input{conclusion.tex}

\section*{Acknowledgment}
\input{ack.tex}

\bibliographystyle{IEEEtran}
\bibliography{ref.bib}

\end{document}

%% file: abstract.tex
Multiple-input multiple-output (MIMO) is a key for the fifth generation (5G) and beyond wireless communication systems owing to higher spectrum efficiency, spatial gains, and energy efficiency. Reaping the benefits of MIMO transmission can be fully harnessed if the channel state information (CSI) is available at the transmitter side. However, the acquisition of transmitter side CSI entails many challenges. In this paper, we propose a deep learning assisted CSI estimation technique in highly mobile vehicular networks, based on the fact that the propagation environment (scatterers, reflectors) is almost identical thereby allowing  a data driven deep neural network (DNN) to learn the non-linear CSI relations with negligible overhead. Moreover, we formulate and solve a dynamic network slicing based resource allocation problem for vehicular user equipments (VUEs) requesting  enhanced mobile broadband (eMBB) and ultra-reliable low latency (URLLC) traffic slices. The formulation considers a threshold rate violation probability minimization for the eMBB slice while satisfying a probabilistic threshold rate criterion for the URLLC slice. Simulation result shows that an overhead reduction of 50\% can be achieved with 12\% increase in threshold violations compared to an ideal case with perfect CSI knowledge. 

%% file: introduction.tex
The exponential growth of smart devices and plethora of emerging applications in 5G and beyond require highly efficient transmission techniques. Initial 5G use cases are expected to support enhanced mobile broadband (eMBB) and ultra reliable low latency communication (URLLC) \cite{ngmn}. While the former requires high peak data rates, the latter focuses on low-latency transmissions with high reliability.  

Vehicle-to-everything (V2X) communication is instrumental in  enabling safer transportation. V2X use cases range from platooning, autonomous driving, collision avoidance, and infotainment services \cite{car}. The mission critical applications of autonomous vehicles require highly reliable and low latency communication. On the other hand, provisioning of entertainment services falls under the eMBB use case, centered on Gbps data rate transmission \cite{Metis}.

Channel state information (CSI) is of paramount importance to maximize the performance of 5G systems relying on spatial diversity harnessing multiple-input multiple-output (MIMO) transmission.
The knowledge of CSI enables designing efficient transmission, scheduling, and user-association schemes. CSI brings extra degree of freedom (DoF) by providing spatial multiplexing and robustness against channel impairments. 
Moreover, 5G systems are expected to utilize higher frequency bands in the millimeter wave (mmWave) range, where communication is contingent upon successful beam alignment.
However, timely and accurate acquisition of CSI is a major hurdle and it is becoming more challenging with the next generation of mobile systems. Acquisition of CSI entails radio resource overhead either in the uplink or downlink whose overhead scales with the channel DoF \cite{csi_acquisition}. 

\subsection{Related Work}
The conventional method of CSI overhead reduction involves the exploitation of linear structures in frequency, spatial, or time domains.
The key idea of overhead reduction lies in determining signal sparsity or exploiting linear correlation in the transform domain i.e., angular domain \cite{csi2}-\cite{csi3}, time domain \cite{csi4}, and frequency domain \cite{csi5}. Furthermore, CSI can also harness non-linear structures (CSI relation between geographically separated users), which are studied using a data driven approach \cite{csi1}-\cite{csi}. The aforementioned deep learning works achieve an overhead reduction by exploiting the non-linear structures of CSIs at the basestation and perform CSI inference at a remote basestation assuming the network has common scatterers. One common limitation of the linear CSI inference is its reliance on instantaneous CSI in the transform domain and the non-linear CSI overhead reduction work involves the CSI estimation at a control basestation. In the real world, obtaining accurate CSI is challenging due to channel estimation errors, thereby inevitably leading to system performance degradation. Moreover, estimating the CSI becomes increasingly difficult when considering mobility. 
On the other hand, \cite{Park_2019} studies the fundamental blocks of machine learning i.e., neural network architectures, training and inference operation, and communication scenarios for several use cases pertaining to various mission critical applications in 5G and beyond. Similarly, several works focus on the resource management of URLLC \cite{Samarakoon_2018}, \cite{Abdel_Aziz_2019}, and eMBB slices \cite{Alsenwi_2019}. The joint design of transmission power and resource allocation for URLLC traffic in vehicular network with distributed learning is studied in \cite{Samarakoon_2018}, and an active learning approach for maximizing the knowledge of network dynamics is presented in \cite{Abdel_Aziz_2019}. Resource allocation of multiple slice network (URLLC, eMBB) with chance constrained optimization is studied in \cite{Alsenwi_2019}. 
Towards this end, the existing resource allocation assume perfect CSI knowledge without considering the inherent challenges of CSI acquisition.  

Therefore in this work, we propose a deep learning based CSI inference framework and present a resource allocation algorithm for mobile vehicular users (VUEs) requesting eMBB and URLLC traffic slices. The main contribution of this paper is as follows:

\begin{itemize}
    \item We introduce a deep learning based CSI inference framework for radio resource overhead reduction in vehicular networks. The data driven based Deep neural network (DNN) learns the non-linear relationship between the CSI of geographically separated VUEs.
    \item We consider a sliced vehicular network with URLLC and eMBB slices and formulate the resource allocation scenario as a rate reliability maximization problem taking into account the error due to CSI inference.
    \item We empirically evaluate the performance of the proposed CSI inference based resource allocation algorithm. 
\end{itemize}

The remainder of the paper is structured as follows. Section II explains the system model and the joint eMBB-URLLC scheduler. CSI inference model and DNN implementation is discussed in Section III. Simulation results are shown in Section IV. Finally, the last section concludes the paper.

%% file: system_model.tex
\subsection{System Model}
\begin{figure}[t]
	\centering
	\includegraphics[trim=2 2 2 2,clip, width=0.5\textwidth]{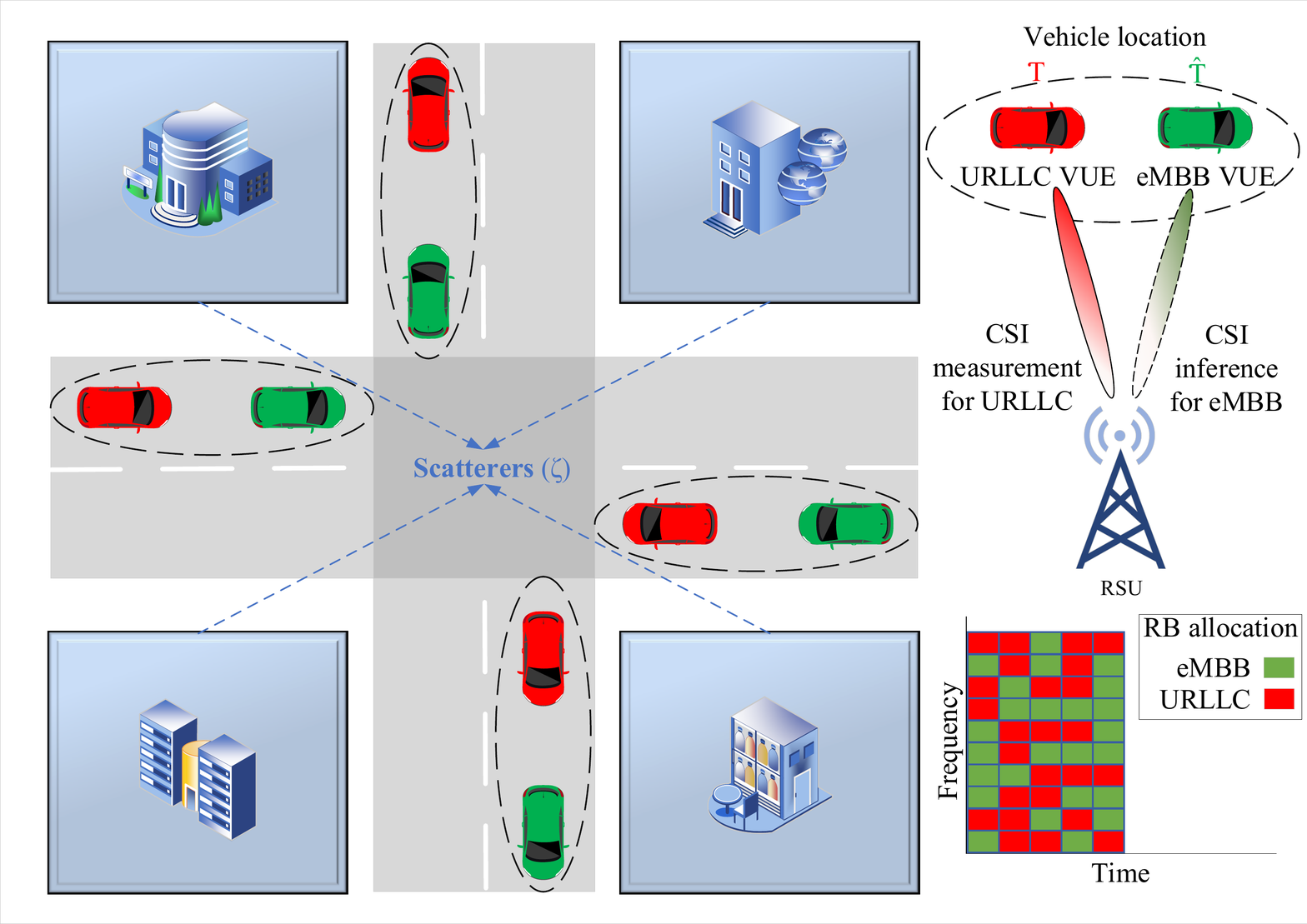}
	\caption{\label{system}Layout of the CSI inference framework.}
\end{figure}

We study a downlink orthogonal frequency-division multiple access (OFDMA) system with MIMO transmission consisting of a set $\mathcal{S}$ of road side units (RSUs) and a set $\mathcal{V}$ of vehicles. 
Vehicles in the network can request the services from the eMBB or the URLLC slice, which share the available bandwidth $B$.
In this view, the set of vehicles are partitioned into $\{\mathcal{V}^{\text{E}}, \mathcal{V}^{\text{U}}\}$, where $\mathcal{V}^{\text{E}}$ represents the vehicles requesting eMBB services and the vehicles requiring URLLC services are denoted as $\mathcal{V}^{\text{U}}$ and it is shown in Fig. \ref{system}. 

We assume that the RSUs are equipped with $N_t$ number of uniform linear array (ULA) antennas, which are spaced at a distance of half wavelength and each vehicle has $N_r$ receive antennas. We utilize the discrete fourier transform (DFT) based codebook, where the size of codebook is equal to the number of transmit antennas. The available bandwidth $B$ at the RSU is divided into resource blocks (RB) and the assignment of RB to a vehicle is denoted with a resource indicator variable $\Omega_{s,v}^b \in \{0,1\}$, where $\Omega_{s,v}^b = 1$ indicates that the RB $b$ of RSU $s$ is allocated to vehicle $v$ and $\Omega_{s,v}^b = 0$ otherwise. 
The received signal by vehicle $v \in \mathcal{V}$ transmitted by RSU $s \in \mathcal{S}$ on RB $b$ at time $t$ is given as:
\begin{equation}
    \boldsymbol{y}_{s,v}^{b,t} = \boldsymbol{h}_{s,v}^{b,t}\boldsymbol{w}_{s,v}^{b,t}\boldsymbol{x}_{s,v}^{b,t} + \boldsymbol{z}_{s,v}^{b,t} + I_{s,v}^{b,t},
\end{equation}
where $\boldsymbol{x}_{s,v}^{b,t}$ is the transmitted data, $\boldsymbol{h}_{s,v}^{b,t} \in \mathbbm{C}^{N_r x N_t}$ is the communication channel from RSU $s$ to vehicle $v$, $\boldsymbol{w}_{s,v}^{b,t} \in \mathbbm{C}^{N_t x N_r}$ is the beamforming vector from the RSU to vehicle, $\boldsymbol{I}_{s,v}^{b,t} = \sum_{s' \in \mathcal{S}/s} \boldsymbol{h}_{s',v}^{b,t}\boldsymbol{w}_{s',v}^{b,t}\boldsymbol{x}_{s',v}^{b,t}$ is the interference from the neighboring RSUs using the same frequency resources, and $\boldsymbol{z}_{s,v}^{b,t} \sim \mathcal{CN}(0,\sigma^2)$ is the additive noise with zero mean and variance $\sigma^2$. 
The wireless downlink channel from the RSU to vehicle is modelled using the geometry-based stochastic channel model (GCSM) \cite{channel} given as:
\begin{equation}
    \label{channel}
    \boldsymbol{h}_{s,v}^\text{LoS} = \frac{\lambda \beta}{2\pi d} \exp\bigg({\frac{-j2\pi d}{\lambda}}\bigg),
\end{equation}
where $d$ is the distance (km) of the direct path between the RSU $s$ and vehicle $v$, $\lambda$ is the wavelength, and $\beta$ corresponds to the complex antenna amplitude. 
The vehicles are associated to the RSU based on maximum received signal strength (max RSSI) using the LoS channel model  \eqref{channel} and the distance dependent path loss model:  $\text{PL}_{dB} = 100.7 + 23.5 \log(d)$ \cite{pathloss}. 
The network is mapped using a system level simulator, which uses the link-to-system (L2S) interface for modelling radio links and require channel state information (CSI) as an input. Maximal ratio combining (MRC) is used at the vehicular devices to exploit spatial diversity and signal-to-interference-plus-noise (SINR) calculations are performed for every RB. Hybrid automatic repeat request (HARQ) mechanism is employed for the re-transmission of failed packets. The achievable rate $R_{v}^t$ by each vehicle depends upon the allocated resources and the downlink beamformer i.e., 

\begin{equation}
    \label{rate}
    R_{v}^t(\boldsymbol{\Omega},\boldsymbol{W}) = \sum_{b=1}^{B} \Omega_{s,v}^{b,t} \omega  \log\big( 1 + \frac{P_{s,v}^{b,t}|\boldsymbol{h}_{s,v}^{b,t}\boldsymbol{w}_{s,v}^{b,t}|^2}{\sigma^{2} + \boldsymbol{I}_{s,v}^{b,t}} \big) ,
\end{equation}
where $\boldsymbol{\Omega}$ represents the resource allocation vector, $\omega$ is the bandwidth of RB, ${\boldsymbol{w}}_{s,v}^{b,t} \in \boldsymbol{W}$ is the transmit beamformer from RSU $s$ to vehicle $v$. 

\subsection{Joint eMBB-URLLC Resource Allocation Policy}
The focus of this work is to allocate RBs to serve eMBB and URLLC VUEs. Towards URLLC, the goal is to ensure the services are provided within the maximum allowed delay to achieve the threshold rate $R^\text{U}$, i.e., $\mathbbm{P}(\Bar{R}_v \leq R^{\text{U}})  \leq  \epsilon, \; {v \in \mathcal{V}^{\text{U}}}$. Here, $\Bar{R}_v = \frac{1}{t} \sum_{\tau=1}^tR_v^{\tau}$ is the time average rate of vehicle and $\epsilon$ is very small threshold violation probability. For eMBB service, it is essential to guarantee the threshold rate by adopting best effort threshold violation probability minimization, i.e., to minimize ${\max\mathbbm{P}(\Bar{R}_v \leq R^{\text{E}}), v\in\mathcal{V}^{\text{E}}}$. In this view, we propose a heuristic resource allocation policy presented in Algorithm 1. Therein, RSUs orthogonally allocate RBs among VUEs with the priority towards URLLC and the remaining resources are scheduled over eMBB VUEs.

CSI is an important input parameter for the L2S interface and it is mapped to the corresponding mutual information curves for the computation of error probability. The rationale in the proposed algorithm is that the URLLC VUEs need to ensure a target reliability, while the eMBB VUEs are minimizing the maximum threshold violation probability using the remaining resources.

\begin{algorithm}[hbtp]
 \caption{Joint eMBB, URLLC scheduler}
 \begin{algorithmic}[1]
 \label{optimization_algorithm}
 \renewcommand{\algorithmicrequire}{\textbf{Input:}}
 \renewcommand{\algorithmicensure}{\textbf{Output:}}
 
 \REQUIRE The set of all vehicles $\mathcal{V}^{\text{E}} \cup \mathcal{V}^{\text{U}}$ and reported CSI $\boldsymbol{C}$
 \ENSURE RB allocation $\boldsymbol{\Omega}$
   \FOR {$\text{time} = 1,...,t$}
  \STATE Compute $\mathbbm{P}(\Bar{R}_v \leq {\text{X}}),\,\text{X} \in \{R^\text{E},R^\text{U}\}$.
  \STATE Estimate $\boldsymbol{W}$ using $\boldsymbol{C}$, $\forall\, {v}$. 
  
  \vspace{3pt}
  \STATExx\subgroup{8}{\emph{{\color{SeaGreen} RB allocation}:}}
    \STATE{Compute the error probability per RB by mapping $\boldsymbol{C}$} \STATEx{\pushcode[0.95]to corresponding MI curves.}
    \FOR {$v \in \mathcal{V}^{\text{U}}$}
  \STATE Allocate RBs to ensure $\mathbbm{P}(\Bar{R}_v \leq R^{\text{U}})  \leq  \epsilon$.
  \ENDFOR
   \FOR {$v \in \mathcal{V}^{\text{E}}$}
    \STATE Use the remaining RBs to ${\min\max\mathbbm{P}(\Bar{R}_v \leq R^{\text{E}})}$.
  \ENDFOR
  \STATE Compute $ R_{v}^t$ as per (3).
  \ENDFOR
 \end{algorithmic} 
 \end{algorithm} 

%% file: CSI_inference.tex
In this section, we investigate the non-linear CSI relationship between geographically separated VUEs and propose a CSI inference framework based on DNN. The aim is to enable low CSI overhead wireless connectivity in a sliced vehicular network via resource slicing. To perform radio resource scheduling of eMBB and URLLC slices we need an estimate of the channel statistics. However, acquiring the accurate CSI entails many challenges especially when the users are mobile and the overhead of CSI estimation may deplete the available radio resources. In this regard, we estimate the CSI of URLLC users due to their strict reliability requirement while the CSI of eMBB users are inferred to avoid resource scarcity. 

\begin{figure}[t]
	\centering
	\includegraphics[trim=2 2 2 2,clip, width=0.5\textwidth]{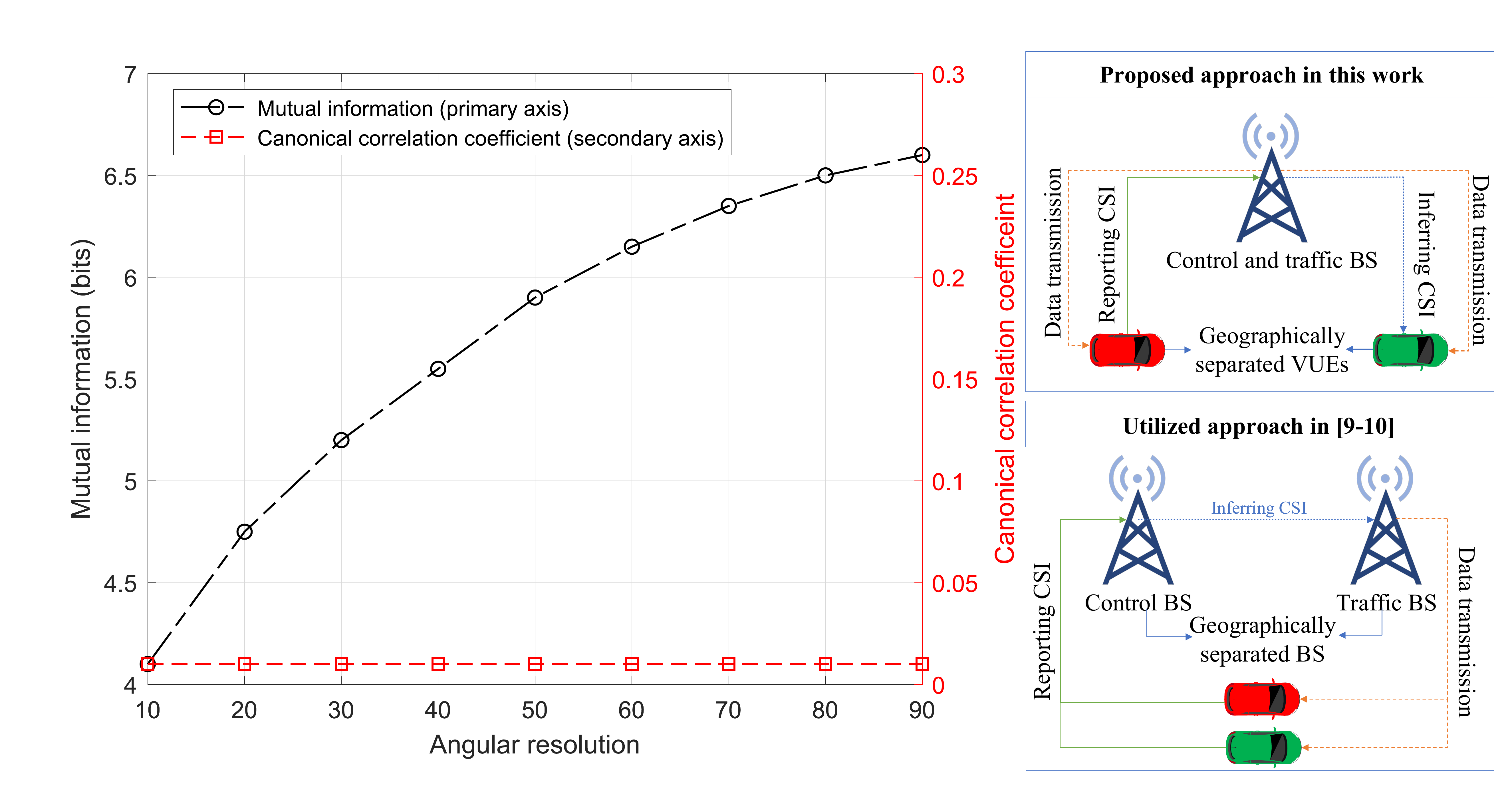}
	\caption{\label{MI} Mutual information and the canonical correlation coefficient of reported and inferred CSI.}
\end{figure}

Towards this end, we start with describing the non-linear relation between the CSI of geographically separated VUEs, which can emerge due to the existence of some location based components or common scatterers as shown in Fig. \ref{system}. It is well known that linear correlations exists among co-located antennas that are of the order of wavelength apart from each other \cite{co-location} and there exists a \emph{region of stationarity} within which the CSI can be treated as a wide sense stationary (WSS) process. To prove the non-linear relationship for geographically separated VUEs, we resort to mutual information (MI) that is a measure of mutual dependency between the two random variables. Fig. \ref{MI} shows the MI and canonical correlation coefficient (measure of the association/correlation among two sets of variables) as a function of angular resolution, where the canonical correlation between the two CSIs is almost zero and the MI is non-zero, hereby validating the point that there exists some non-linear relation. Moreover, the proposed approach is different from the works \cite{csi1}-\cite{csi}, in that we infer the CSI of a neighboring vehicle while these works study the inference problem for a neighboring basestation as shown in Fig. \ref{MI}. The works \cite{csi1}-\cite{csi} infer the CSI of the same user at a remote basestation, while we infer the CSI of remote user for the same basestation.

\subsection{CSI Inference and RB Allocation}
CSI inference of a remote vehicle given the reported CSI of another neighboring vehicle is proposed. CSI at the RSU enable the design of the downlink beamformers, which are used for spatial diversity and multiple spatial access. In particular, we consider a communication system consisting of mobile vehicles which have the common scatterers. Lets consider a URLLC vehicle $v$ whose CSI $\boldsymbol{c}_v$ is reported and corresponding downlink beamformer $\boldsymbol{w}_v$ is estimated by the RSU i.e., $|\boldsymbol{h}_{s,v}^{b,t}\boldsymbol{w}_{s,v}^{b,t}(\boldsymbol{c}_v)|^2 = 1$. Furthermore, let there be a neighboring eMBB vehicle $v'$ whose CSI $\boldsymbol{\hat{c}}_{v'}$ and the corresponding downlink beamformers $\boldsymbol{\hat{w}}_{v'}$ are inferred i.e., $|\boldsymbol{h}_{s,v'}^{b,t}\boldsymbol{\hat{w}}_{s,v'}^{b,t}(\boldsymbol{\hat{c}}_{v'})|^2 \leq 1$. The aim of this work, is to infer $\boldsymbol{\hat{w}}_{v'}(\boldsymbol{\hat{c}}_{v'})$ given the knowledge of $\boldsymbol{{w}}_{v}(\boldsymbol{c}_v)$ to reduce the overhead of CSI estimation in the network. 
Mathematically, we can cast the downlink beamformer inference problem as a function mapping of user location and the environment geometry i.e., obstacles, reflection and attenuation coefficients as follows:  

\begin{equation}
    \label{mapping}
    \boldsymbol{w}_v = f(\boldsymbol{c}_v, \boldsymbol{\Gamma}, \boldsymbol{\zeta}) \hspace{5pt} \text{and} \hspace{5pt} \boldsymbol{\hat{w}}_{v'} = f(\boldsymbol{\hat{c}}_{v'}, \boldsymbol{\hat{\Gamma}}, \boldsymbol{\zeta}),
\end{equation}
where $f(\cdot)$ denotes the function mapping of downlink beamformer, $(\boldsymbol{\Gamma}, \boldsymbol{\hat{\Gamma}})$ denote the terminal locations of URLLC and eMBB vehicle respectively, and $\boldsymbol{\zeta}$ specifies the common propagation environment. 
The beamformer mapping model presented in \eqref{mapping} is indirect and infeasible (since there is no linear relation) for theoretical analysis, therefore we adopt a deep learning based function mapping approach at the expense of large training data. 
Training the NN in offline manner helps to efficiently solve user scheduling and resource allocation problems with less overhead. 
The DNN is trained using the beamforming loss function $L$ defined as \cite{csi}: 
\begin{equation}
    \label{mapping_error}
    L(\boldsymbol{\hat{w}}_{v'},\boldsymbol{\hat{w}}_{v'}^{opt}) = 1 - \frac{\boldsymbol{\hat{w}}_{v'}\boldsymbol{w}_{v'}}{\boldsymbol{\hat{w}}_{v'}^{opt}\boldsymbol{w}_{v'}},
\end{equation}
where $\boldsymbol{\hat{w}}_{v'}$ is the inferred beamformer of vehicle $v'$ from the DFT codebook, $\boldsymbol{\hat{w}}_{v'}^{opt}$ is the optimal beamforming vector in the DFT codebook, and $\boldsymbol{w}_{v'}$ is the beamforming label. After the training has converged to an acceptable loss rate, the low CSI overhead resource allocation follows Algorithm 2.

\begin{algorithm}[hbtp]
 \caption{Joint eMBB, URLLC scheduler with reduced CSI overhead}
 \begin{algorithmic}[1]
 \label{rb_algorithm}
 \renewcommand{\algorithmicrequire}{\textbf{Input:}}
 \renewcommand{\algorithmicensure}{\textbf{Output:}}

  \REQUIRE The set of URLLC vehicles $\mathcal{V}^{\text{U}}$ and their reported CSI 
  \STATEx{\pushcode[1.65]$\boldsymbol{c_v}, \forall v \in \mathcal{V}^{\text{U}}$}
  \ENSURE RB allocation $\boldsymbol{\Omega}$
  
  \FOR {$\text{time} = 1,...,t$}
  \STATE Compute $\mathbbm{P}(\Bar{R}_v \leq {\text{X}}),\,\text{X} \in \{R^\text{E},R^\text{U}\}$.
  \STATE Estimate $\boldsymbol{w}_v$ using $\boldsymbol{c}_v$, $\forall\, {v} \in \mathcal{V}^{\text{U}}$.

  \vspace{3pt}
  \STATExx\subgroup{5}{\emph{{\color{SeaGreen} RB allocation for URLLC}:}}
    \STATE{Compute the error probability per RB by mapping \\
    \pushcode[0.95] $ \boldsymbol{c_v},  \forall v \in \mathcal{V}^{\text{U}}$ to corresponding MI curves.}
  \FOR {$v \in \mathcal{V}^{\text{U}}$}
   \STATE Allocate RBs to ensure $\mathbbm{P}(\Bar{R}_v \leq R^{\text{U}})  \leq  \epsilon$.
  \ENDFOR
  
  \renewcommand{\algorithmicrequire}{\textbf{\hspace{23pt} DNN input:}}
  \renewcommand{\algorithmicensure}{\textbf{\hspace{23pt} DNN output:}}

    \vspace{3pt}

   \STATExx\subgroup{4}{\emph{{\color{NavyBlue} CSI inference for eMBB}:}}
  \REQUIRE $\boldsymbol{c_v},\,v\in \mathcal{V}^{\text{U}}$
  \STATE{\hspace{8pt} Infer CSI and beamformer of eMBB vehicles via}
  \STATEx{\pushcode[2.1]DNN.}
  \ENSURE $\boldsymbol{c_{v'}}, \boldsymbol{w_{v'}},\,v'\in \mathcal{V}^{\text{E}}$

  \vspace{3pt}
  \STATExx\subgroup{5}{\emph{{\color{SeaGreen} RB allocation for eMBB}:}}
  \STATE{Compute the error probability for remaining RBs by} 
  \STATEx{\pushcode[0.95] mapping $\boldsymbol{c_{v'}}, \,v'\in \mathcal{V}^{\text{E}}$ to corresponding MI curves.}
  \FOR {$v' \in \mathcal{V}^{\text{E}}$}
      \STATE Use the remaining RBs to ${\min\max\mathbbm{P}(\Bar{R}_{v'} \leq R^{\text{E}})}$.
  \ENDFOR
  
  \STATE  Compute $ R_{v}^t$ as per (3).
  \ENDFOR
 \end{algorithmic} 
 \end{algorithm} 

\begin{figure}[t]
	\centering
	\includegraphics[trim=4 4 4 4,clip, width=0.5\textwidth]{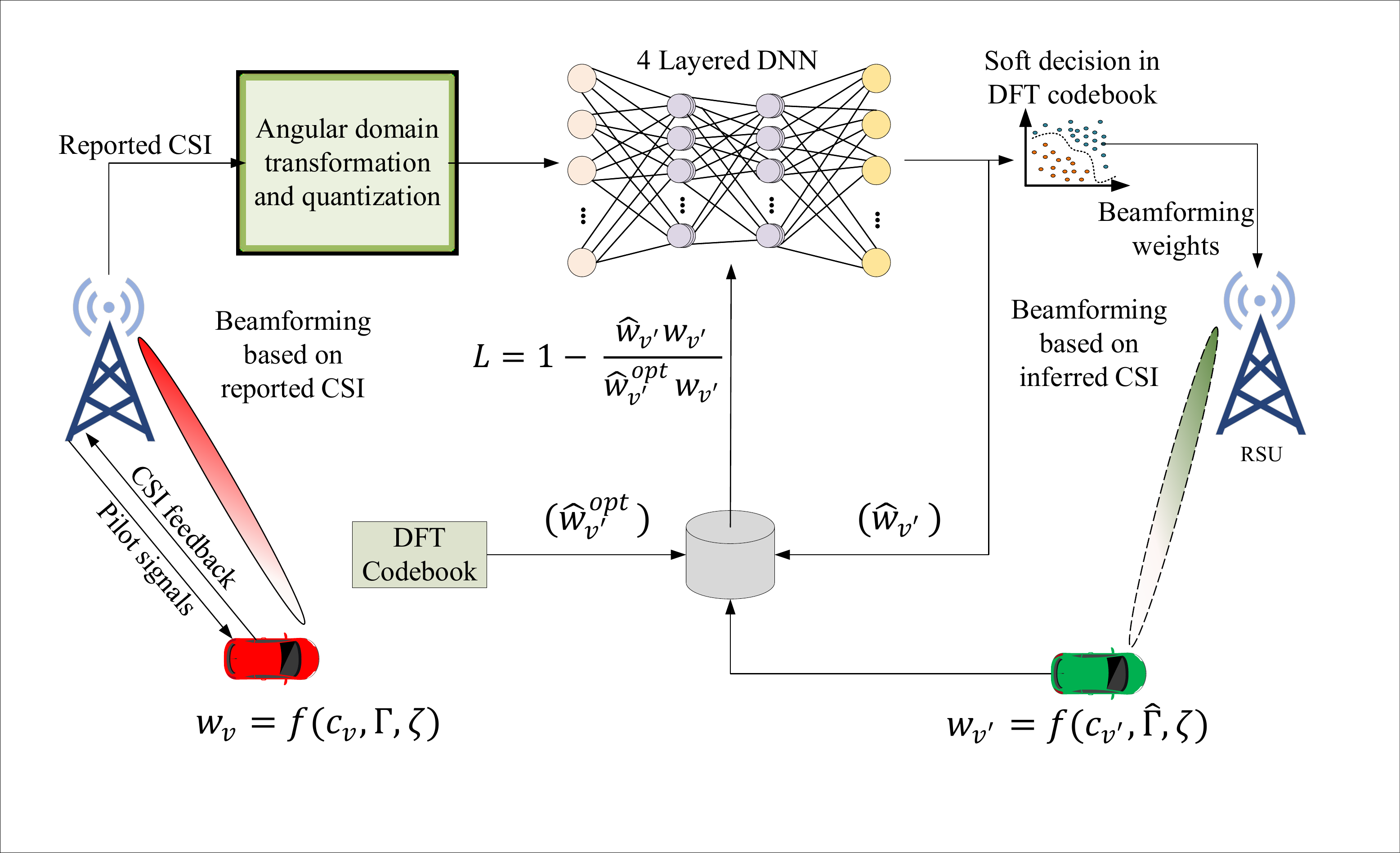}
	\caption{\label{dnn} DNN architecture for remote beamforming inference.}
\end{figure}

\subsection{DNN Implementation}
The non-linear CSI structure exploitation problem is solved with deep neural networks (DNNs) based on a data-driven approach, since it does not require any explicit model of the propagation environment. 
In our work, the training data for the DNN consist of the CSI of the geographically separated users which is obtained through pilot aided channel estimation in the offline training phase. The input data is further transformed into the angular domain followed by quantization to enable feature extraction, which represents the large-scale fading properties of the wireless channel \cite{sparse}. The CSI angular domain representation exhibits sparsity which improves the training of the DNN by lowering the inference error in \eqref{mapping_error}. Once the offline training has converged to an acceptable inference error, we proceed with the prediction phase. The low complexity online prediction phase uses the estimated beamformer of URLLC vehicle and infers the beamforming vector of the neighboring eMBB vehicle based on the learned policy. In the online phase the output of the DNN is a soft decision which indicates the probability of downlink beamformers $\boldsymbol{\hat{w}}_{v'}(\boldsymbol{\hat{c}}_{v'})$ in the DFT codebook. In this way, the non-linear CSI exploitation helps to enhance the performance of cellular network by reducing the overhead of CSI estimation i.e., given the estimate of beamformer of one user we can infer the beamformer of another. The inference problem per vehicle and per training sample can be mathematically written as:

\begin{align}
\label{dnn_error}
\begin{array}{rrclcl}
& \text{Given} & \multicolumn{2}{l}{\boldsymbol{{w}}(\boldsymbol{{c}}_{v}), v \in \mathcal{V}^{\text{U}}} &\; & \\
& \displaystyle\min & \multicolumn{2}{l}{L(\boldsymbol{\hat{w}}_{v'},\boldsymbol{\hat{w}}_{v'}^{opt})}  &\; & \\
\end{array}
\end{align}

\begin{figure*}[htp]
  \centering
  \subfigure[Sparse vehicular network]{\includegraphics[width=0.49\textwidth]{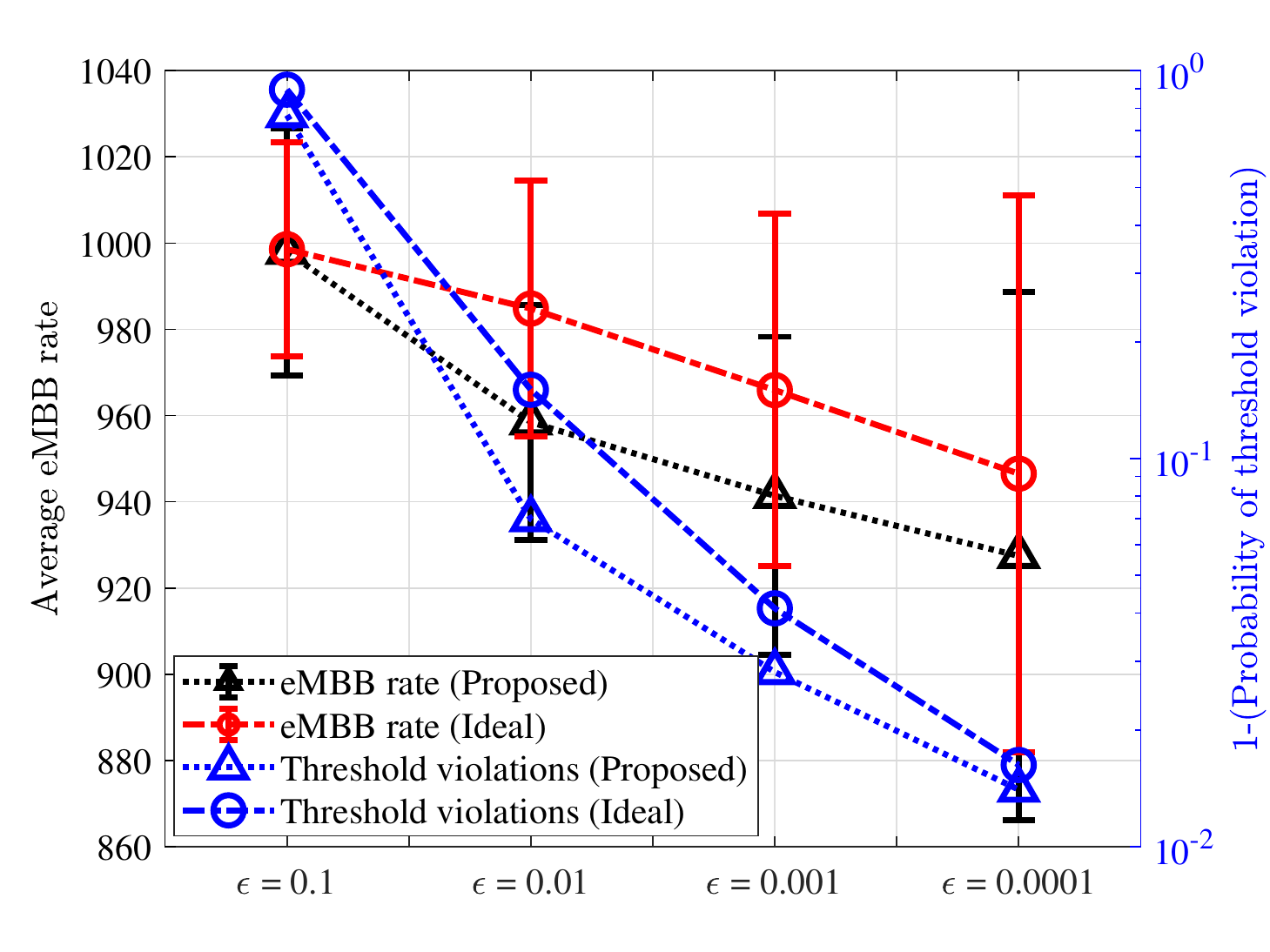}}\quad
  \subfigure[Dense vehicular network]{\includegraphics[width=0.49\textwidth]{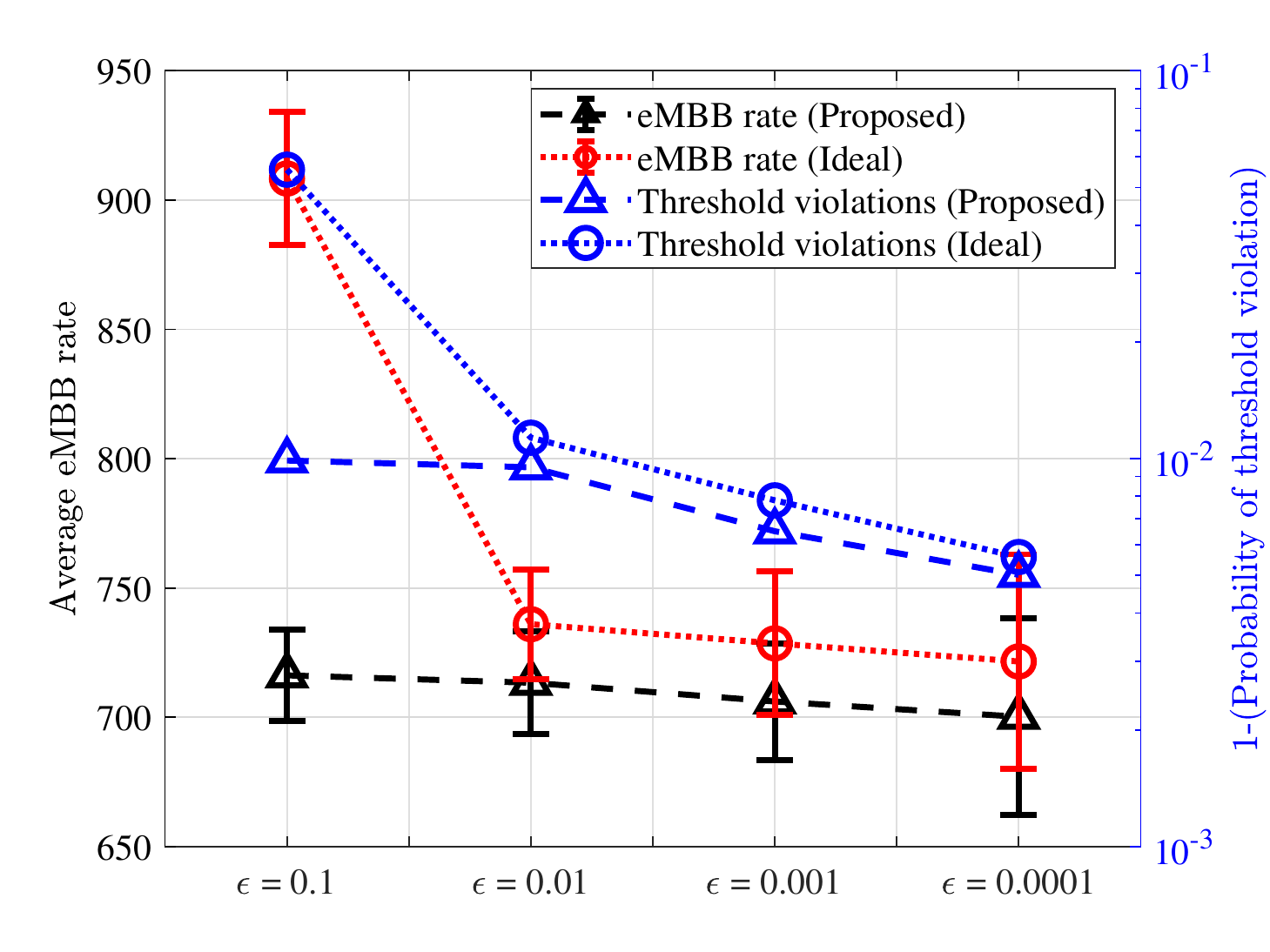}}
  \caption{\label{epsilon}Effect of reliability ($\epsilon$) on the average eMBB rate and the threshold rate violations.}
\end{figure*}

The DNN deployed in the RSU utilizes four fully connected layers as shown in Fig. \ref{dnn}. All the layers use rectified linear activation units (ReLU) activation except the output layer. The transformed and quantized CSI data from input layer is aggregated at a fully connected hidden layer that is composed of 128 neurons and the aggregated data is passed on to the output layer, which applies a softmax activation function. The training of the DNN uses Adam optimizer \cite{adam} and rest of the DNN parameters are listed in Table. \ref{sim_param}. 

\begin{table}[t]
\caption{Simulator parameters.}
\centering
\label{sim_param}
\begin{tabular}{l l}
\hline
\textbf{Parameter}   & \textbf{Value}             \\ \hline \hline
Scenario             & 3GPP 3D Urban micro  \\
Bandwidth  & 10\,MHz (50 PRB)               \\ 
Time slot (TTI)             & 1\,ms                      \\ 
Carrier frequency          & 5.9\,GHz \\ 
L2S interface         & MIESM                     \\
Traffic   & Periodic                          \\
Tx Power             & 30\,dBm                     \\ 
Tx antenna elements  & 64                        \\ 
Rx antenna elements  & 8                        \\ 
Target rate URLLC ($R^{\text{U}}$) & 128 kbps \\
Target rate eMBB ($R^{\text{E}}$) & 1000 kbps \\
\hline
\hline \multicolumn{2}{c}{ \textbf{DNN Parameters} } \\ \hline
Number of layers & 4 \\
Learning rate & 0.0001 \\
Discount factor & 0.99 \\
Number of neurons per layer & 128 \\
Input & Transformed and quantized CSI \\
Output & Soft decision of CSI in DFT codebook \\
\hline
\end{tabular}
\end{table}

%% file: simulation.tex
In this section we present the simulations results and evaluate the performance of the proposed algorithm. The proposed eMBB, URLLC scheduler is evaluated in a vehicular network, where the vehicles are moving with a speed of 40\,km/h and the rest of the simulation parameters are tabulated in Table \ref{sim_param}.  

The performance of the proposed resource allocation algorithm is analyzed in two scenarios (i) \emph{Sparse} network (the resources in the network are more than the vehicles) and (ii) \emph{Dense} network (the resources are less than the vehicles), which are represented in Fig. \ref{epsilon}. The density of vehicles is varied by changing the inter-vehicular distance. The proposed technique reflects the case when CSI is reported by the URLLC users and the CSI of eMBB vehicles is inferred. The performance of the proposed technique is compared with the ideal case, which represents the scenario with perfect CSI knowledge i.e., CSI is reported by both URLLC and eMBB vehicles. 

The average eMBB rates (left y-axis) and the threshold satisfaction probability (right y-axis) as a function of vehicular density, and URLLC reliability ($\epsilon$) is shown in Fig. \ref{epsilon}, while $R^{\text{U}} = 128\,\text{kbps}$,\, $R^{\text{E}} = 1000\,\text{kbps}$ is fixed. The proposed eMBB-URLLC scheduler is aimed to satisfy the URLLC rate constraint with $\epsilon$ reliability and afterwards use the best effort approach to satisfy the eMBB target rate. In case of the sparse network, where the vehicular density is less than the available resources in the network we can see that eMBB rate target is achieved with a higher probability for lesser URLLC reliability values. On the other hand, when the URLLC reliability is increased ($\epsilon = 0.001, 0.0001$) there is a decrease in vehicles satisfying the eMBB target rate. The vehicles in the sparse case satisfy the eMBB target rate with $76.8\%$ probability when the reliability of URLLC users is $\epsilon = 0.1$ compared to the ideal scenario which satisfies the threshold rate with $89\%$ probability. The probability of users satisfying the eMBB target for the proposed technique decreases to $6.9\%$, when the reliability is increased to $\epsilon = 0.01$ and it decreases further with the increasing reliability i.e, the target rate is achieved by less than $1.4\,\%$ of the vehicles with maximum reliability $\epsilon = 0.0001$. Furthermore, when the number of vehicles in the network is increased while keeping the resource same, we see a decrease in the number of users satisfying the eMBB target rate. The dense network with reliability $\epsilon = 0.1$ achieves the target rate for $0.9$\,\% of the vehicles and it is reduced to $0.5$\,\% when the reliability is increased to $\epsilon = 0.0001$. On the other hand, ensuring the reliability of the URLLC vehicles is related to the resource utilization (e.g., parity, redundancy, and re-transmissions) \cite{mehdi} i.e., lower reliability results into lower resource consumption and vice-versa. From Fig. \ref{epsilon} we can see that as the reliability is increased the eMBB rate decreases which is due to the lesser number of available resources. Moreover, the standard deviation of eMBB rates increases with the increase in reliability i.e., higher rate fluctuations are observed for high reliability. 

\begin{figure}[t]
	\centering
	\includegraphics[width=0.5\textwidth]{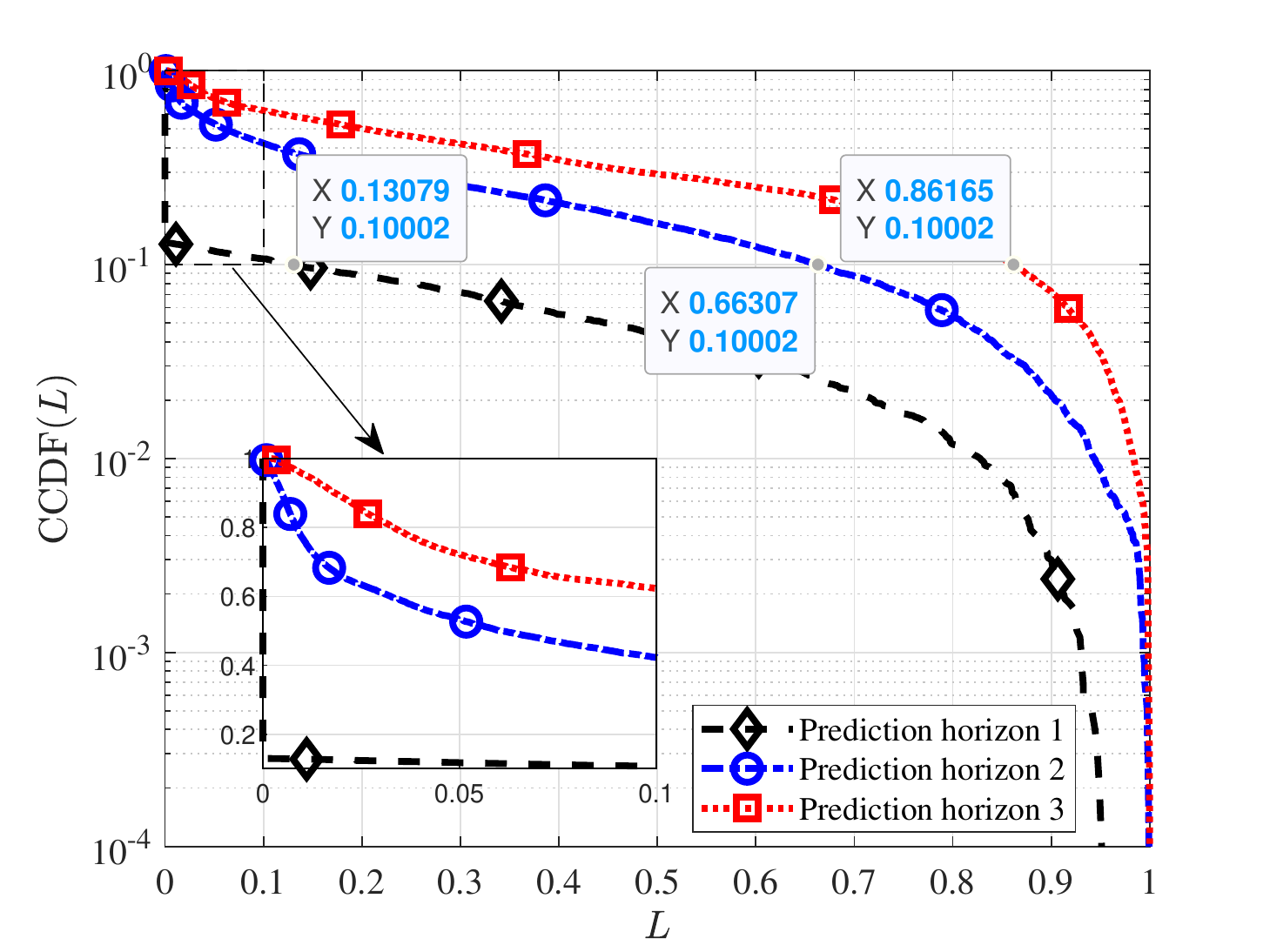}
	\caption{\label{beamforming} CCDF of the beamforming loss of inference for different prediction horizon.}
	\vspace{-5pt}
\end{figure}

The performance of the CSI inference framework is presented in Fig. \ref{beamforming}, which shows the effect of CSI inference on the beamforming loss. The complementary cumulative distribution function (CCDF) illustrates the beamforming loss during inference, where prediction horizon 1 refers to the inference of same time index $\{t\}$, prediction horizon 2 refers to time indices $\{t,\,t+1\}$, and so on.  
The observed beamforming loss for $90\,\%$ of the inference for prediction horizon 1 is $0.13$, and the loss increases to $0.66$ when the inference is performed for prediction horizon 2, and it becomes worse as we try to infer further in the future. Moreover, it is shown in Fig. \ref{beamforming} that optimal beamformers can be inferred with $L = 0$ for $87\%$ of the inference in case of prediction horizon 1, and for prediction horizon 3 the optimal beamformers will always be inferred with $L > 0$. The beamforming loss increases as we try to infer in the future, which is due to the fact that mobile VUEs will have dynamic propagation environment making the channels highly fluctuating and making error-free predictions extremely unlikely. 

%% file: conclusion.tex
In this paper, we have formulated a low CSI overhead based eMBB, URLLC scheduler, which satisfies the URLLC constraints and minimizes the threshold rate violation probability of eMBB users.
We considered a scenario of multiple serving RSUs, vehicles, and evaluated the performance of the proposed scheduler under different URLLC reliability values, density of vehicles, and multiple inference prediction horizon. 
The CSI inference problem considered in this work is solved using a DNN, which is capable of mapping the non-linear relationship for the CSI exploitation problem.
The existence of non-linear relationships between the CSIs of geographically separated users is characterized using mutual information. 
The CCDF of the loss function shows that CSI can be inferred for prediction horizon 1 with 0.13 inference error for almost 90\,\% of the times and the CSI inference results into 12\% increase in threshold violations. 
The proposed approach considerably reduces the CSI overhead of eMBB vehicles due to the existence of non-linear CSI relationship.

%% file: ack.tex
This work is supported by the projects SAFARI, High5, and 6Genesis Flagship. First author is grateful for the grants received from HPY, Nokia, and Walter Ahlströmin Foundation.

%% file: main.bbl
\begin{thebibliography}{10}
\providecommand{\url}[1]{#1}
\csname url@samestyle\endcsname
\providecommand{\newblock}{\relax}
\providecommand{\bibinfo}[2]{#2}
\providecommand{\BIBentrySTDinterwordspacing}{\spaceskip=0pt\relax}
\providecommand{\BIBentryALTinterwordstretchfactor}{4}
\providecommand{\BIBentryALTinterwordspacing}{\spaceskip=\fontdimen2\font plus
\BIBentryALTinterwordstretchfactor\fontdimen3\font minus
  \fontdimen4\font\relax}
\providecommand{\BIBforeignlanguage}[2]{{%
\expandafter\ifx\csname l@#1\endcsname\relax
\typeout{** WARNING: IEEEtran.bst: No hyphenation pattern has been}%
\typeout{** loaded for the language `#1'. Using the pattern for}%
\typeout{** the default language instead.}%
\else
\language=\csname l@#1\endcsname
\fi
#2}}
\providecommand{\BIBdecl}{\relax}
\BIBdecl

\bibitem{ngmn}
\BIBentryALTinterwordspacing
NGMN, ``{5 x 5G Five things you need to know about 5G and what it delivers},''
  2018. [Online]. Available:
  \url{https://www.ngmn.org/fileadmin/ngmn/content/downloads/Technical/2018/180120_NGMN_Value_Proposition_on_5G.pdf}
\BIBentrySTDinterwordspacing

\bibitem{car}
A.~E. Fernandez and M.~Fallgren, ``{5GCAR} scenarios, use cases, requirements
  and kpis,'' Fifth Generation Communication Automotive Research and
  Innovation, 2017.

\bibitem{Metis}
\BIBentryALTinterwordspacing
``{METIS deliverable D7.3 Final 5G visualization},'' 2017. [Online]. Available:
  \url{https://metis-ii.5g-ppp.eu/wp-content/uploads/deliverables/METIS-II_D7.3_V1.0.pdf}
\BIBentrySTDinterwordspacing

\bibitem{csi_acquisition}
Z.~{Jiang}, A.~F. {Molisch}, G.~{Caire}, and Z.~{Niu}, ``{On the achievable
  rates of FDD massive MIMO systems with spatial channel correlation},'' in
  \emph{2014 IEEE/CIC International Conference on Communications in China
  (ICCC)}, Oct 2014, pp. 276--280.

\bibitem{csi2}
J.~{Brady}, N.~{Behdad}, and A.~M. {Sayeed}, ``{Beamspace {MIMO} for
  Millimeter-Wave Communications: System Architecture, Modeling, Analysis, and
  Measurements},'' \emph{IEEE Transactions on Antennas and Propagation},
  vol.~61, no.~7, pp. 3814--3827, Jul 2013.

\bibitem{csi3}
X.~{Rao} and V.~K.~N. {Lau}, ``{Distributed Compressive {CSIT} Estimation and
  Feedback for {FDD} Multi-User Massive {MIMO} Systems},'' \emph{IEEE
  Transactions on Signal Processing}, vol.~62, no.~12, pp. 3261--3271, Jun
  2014.

\bibitem{csi4}
R.~{Hadani}, S.~{Rakib}, M.~{Tsatsanis}, A.~{Monk}, A.~J. {Goldsmith}, A.~F.
  {Molisch}, and R.~{Calderbank}, ``{Orthogonal Time Frequency Space
  Modulation},'' in \emph{2017 IEEE Wireless Communications and Networking
  Conference (WCNC)}, Mar 2017, pp. 1--6.

\bibitem{csi5}
L.~{You}, X.~{Gao}, A.~L. {Swindlehurst}, and W.~{Zhong}, ``{Channel
  Acquisition for Massive {MIMO-OFDM} With Adjustable Phase Shift Pilots},''
  \emph{IEEE Transactions on Signal Processing}, vol.~64, no.~6, pp.
  1461--1476, 2016.

\bibitem{csi1}
Z.~Jiang, S.~Chen, A.~F. Molisch, R.~Vannithamby, S.~Zhou, and Z.~Niu,
  ``{Exploiting Wireless Channel State Information Structures Beyond Linear
  Correlations: A Deep Learning Approach},'' \emph{IEEE Communications
  Magazine}, vol.~57, no.~3, pp. 28--34, 2019.

\bibitem{csi}
Z.~Jiang, Z.~He, S.~Chen, A.~F. Molisch, S.~Zhou, and Z.~Niu, ``{Inferring
  remote channel state information: Cram{\'e}r-Rae lower bound and deep
  learning implementation},'' in \emph{2018 IEEE Global Communications
  Conference (GLOBECOM)}.\hskip 1em plus 0.5em minus 0.4em\relax IEEE, 2018,
  pp. 1--7.

\bibitem{Park_2019}
\BIBentryALTinterwordspacing
J.~Park, S.~Samarakoon, M.~Bennis, and M.~Debbah, ``{Wireless Network
  Intelligence at the Edge},'' \emph{Proceedings of the IEEE}, vol. 107,
  no.~11, p. 2204-2239, Nov 2019. [Online]. Available:
  \url{http://dx.doi.org/10.1109/JPROC.2019.2941458}
\BIBentrySTDinterwordspacing

\bibitem{Samarakoon_2018}
\BIBentryALTinterwordspacing
S.~Samarakoon, M.~Bennis, W.~Saad, and M.~Debbah, ``{Federated Learning for
  Ultra-Reliable Low-Latency V2V Communications},'' \emph{2018 IEEE Global
  Communications Conference (GLOBECOM)}, Dec 2018. [Online]. Available:
  \url{http://dx.doi.org/10.1109/glocom.2018.8647927}
\BIBentrySTDinterwordspacing

\bibitem{Abdel_Aziz_2019}
\BIBentryALTinterwordspacing
M.~K. Abdel-Aziz, S.~Samarakoon, M.~Bennis, and W.~Saad, ``{Ultra-Reliable and
  Low-Latency Vehicular Communication: An Active Learning Approach},''
  \emph{IEEE Communications Letters}, p. 1 1, 2019. [Online]. Available:
  \url{http://dx.doi.org/10.1109/LCOMM.2019.2956929}
\BIBentrySTDinterwordspacing

\bibitem{Alsenwi_2019}
\BIBentryALTinterwordspacing
M.~Alsenwi, N.~H. Tran, M.~Bennis, A.~Kumar~Bairagi, and C.~S. Hong,
  ``{eMBB-URLLC Resource Slicing: A Risk-Sensitive Approach},'' \emph{IEEE
  Communications Letters}, vol.~23, no.~4, p. 740 743, Apr 2019. [Online].
  Available: \url{http://dx.doi.org/10.1109/LCOMM.2019.2900044}
\BIBentrySTDinterwordspacing

\bibitem{channel}
P.~Kyasti, J.~Meinila, L.~Hentila, X.~Zhao, T.~Jamsa, C.~Schneider,
  M.~Narandzic, M.~Milojevia, A.~Hong, J.~Ylitalo, V.-M. Holappa,
  M.~Alatossava, R.~Bultitude, Y.~Jong, and T.~Rautiainen, ``{WINNER {II}
  channel models},'' \emph{IST-4-027756 WINNER II D1.1.2 V1.2}, Feb 2008.

\bibitem{pathloss}
3GPP, ``{{TR} 36.814 V9.0.0: Further advancements for {E-UTRA} physical layer
  spects (Release 9)},'' Mar 2010.

\bibitem{co-location}
A.~{Adhikary}, J.~{Nam}, J.~{Ahn}, and G.~{Caire}, ``{Joint Spatial Division and Multiplexing The Large-Scale Array Regime},'' \emph{IEEE Transactions
  on Information Theory}, vol.~59, no.~10, pp. 6441 6463, Oct 2013.

\bibitem{sparse}
C.~Studer, S.~Medjkouh, E.~G{\"o}n{\"u}lta{\c{s}}, T.~Goldstein, and
  O.~Tirkkonen, ``{Channel charting: Locating users within the radio
  environment using channel state information},'' \emph{IEEE Access}, vol.~6,
  pp. 47\,682--47\,698, 2018.

\bibitem{adam}
D.~Kingma and J.~Ba, ``{Adam: A Method for Stochastic Optimization},''
  \emph{International Conference on Learning Representations}, Dec 2014.

\bibitem{mehdi}
M.~{Bennis}, M.~{Debbah}, and H.~V. {Poor}, ``{Ultrareliable and Low-Latency
  Wireless Communication: Tail, Risk, and Scale},'' \emph{Proceedings of the
  IEEE}, vol. 106, no.~10, pp. 1834--1853, Oct 2018.

\end{thebibliography}
